\documentstyle[aps,twocolumn]{revtex}

\def\be{\begin{equation}}
\def\ee{\end{equation}}
\def\disp{\displaystyle}

\newcounter{fig}

\begin{document}
\thispagestyle{empty}

\title{Molecular Weight Dependence of 
Spreading Rates of Ultrathin Polymeric Films.}

\author{M.P.Valignat$^1$, G.Oshanin$^{2,3}$, S.Villette$^1$,  A.M.Cazabat$^1$
and M.Moreau$^2$}

\address{$^1$  Laboratoire de Physique de la Mati\`ere Condens\'ee,\\
Coll\`ege de France, 11 place M.Berthelot, 75252 Paris Cedex 05, France}

\address{$^2$  Laboratoire de Physique Th\'eorique des Liquides,\\
Universit\'e Paris VI, 4 place Jussieu, 75252 Paris Cedex 05, France}

\address{$^3$ Centre de Recherche en Mod\'elisation Mol\'eculaire,\\
Universit\'e de Mons-Hainaut, 20 place du parc, 7000 Mons, Belgium}

\address{\rm (Received: December 11, 1997)}
\address{\mbox{ }}
\address{\parbox{14cm}{\rm \mbox{ }\mbox{ }
We study experimentally the molecular weight $M$ dependence of
spreading rates 
of molecularly 
thin precursor films, growing at the bottom of droplets of
polymer liquids.   
In accord with previous observations, we find  that
the radial extension $R(t)$ of the film grows with time as 
$R(t) = (D_{exp} t)^{1/2}$. 
Our data substantiate the $M$-dependence of $D_{exp}$; we
show that it follows $D_{exp} \sim M^{-\gamma}$,
where  
the exponent
$\gamma$
is dependent on
the chemical composition
 of the solid surface, determining its 
frictional properties with respect to the molecular transport.
In the specific case of hydrophilic substrates, 
the frictional properties can be modified by 
the change of the relative humidity
(RH). We find that $\gamma \approx 1$  at 
low RH and tends to zero when RH gets progressively increased.
We propose simple theoretical arguments
 which explain the
observed behavior in 
the limits of low and high RH.
}}
\address{\mbox{ }}
\address{\parbox{14cm}{\rm PACS No: 68.45.-v, 68.10.Gw, 68.45.G, 83.10.Nn}}
\maketitle

\makeatletter
\global\@specialpagefalse

\makeatother

\pagebreak

Experimental studies of 
non-volatile drops spreading
 spontaneously on solid surfaces
often reveal thin precursor films \cite{leg,caza}. 
For some liquid/solid systems
only a single well-defined
precursor film develops from 
the macroscopic droplet. 
In other cases, 
the films
have a remarkable
"terraced" shape being 
formed by  
superimposed compact monolayers 
\cite{caza}. 
For liquids composed of polymers with low molecular weight $M$, 
(i.e. such that $M$ is well below  the entanglement threshold $M_{c}$
in $3D$ \cite{pggb}),
the experimentally observed thickness of 
the precursor film or of the "terraces" is typically
 of order of several angstroms. 
It is
much less than the gyration radius $R_{g}$
 of the same polymers in solution 
  and  corresponds to the
transverse size of the monomeric units. 
This implies that
contrary to the behavior expected for
polymers with high $M$, $M > M_{c}$, for which
 the predicted thickness of
 the precursor film should compare to 
$R_{g}$ \cite{joanny},
 for low-$M$ polymer liquids  
the macromolecules 
in the precursor film 
(or in the superimposed monolayers) are essentially disentangled
and lay 
flat on the solid substrate or on the lower layers.

Most of available experimental studies 
were devoted to the analysis of the temporal evolution
of  precursor films
\cite{leg,caza}. Meanwhile, there is an ample 
evidence \cite{caza,fraysse,vil} that at sufficiently short times,
when the macroscopic droplet still acts as a reservoir,
the radial extension $R(t)$
of such films
follows a universal time-dependence of the form
\begin{equation}
R(t) \; = \; \sqrt{D_{exp} t}
\end{equation}
In view of the form of Eq.(1) the prefactor $D_{exp}$ 
 is often referred to as the "diffusion coefficient" of the precursor.
Needless to say that  
$D_{exp}$ is
different from 
the conventional
diffusion coefficient 
describing particles 
random motion in the bulk 
liquid phase or
on solid substrates; as a matter of fact, 
$D_{exp}$ also depends on
the driving forces which cause the film
spreading. 
 
Physical processes underlying the temporal behavior
as in Eq.(1) are now rather well understood. Moreover, 
several theoretical models which explain
the $\sqrt{t}$-dependence of $R(t)$ 
are now available \cite{dgc,bur}.
However, still little is known about
the
dependence of $D_{exp}$
 on different physical and chemical factors, such as, for example, 
the temperature, relative humidity (RH), chemical composition of the substrate
 and, for polymer liquids, the molecular
weight 
of polymers. On the other hand, 
especially the latter point is 
of essential
importance for many practical applications, such as, e.g. coating, gluing 
or lubrication,  - 
most of liquids used in these material processing operations
are either polymer 
liquids or contain
polymeric additives.

A few available by now studies
of polymeric films spreading on solid surfaces 
concern the molecular weight dependence of $D_{exp}$. Early experimental analysis
 \cite{novotny}
of spreading rates of mesoscopically thin films of perfluorocarbon polymers
suggest that 
 $D_{exp}$ decreases algebraically 
with an increase of $M$, i.e. $D_{exp}$ obeys
\begin{equation}
D_{exp} \; \sim \; \frac{1}{M^{\gamma}},
\end{equation}
where the experimentally observed value of the 
exponent $\gamma$ is close to $1.7$. 
The same behavior was recently recovered
in \cite{marchon} (see also \cite{yoon}) for
low-$M$ perfluoropolyalkylether
polymers, used in the computer 
 industry for lubricating  the hard 
discs in order to reduce wear. 
Studies of 
low-$M$ polydimethylsiloxane 
(hereafter abbreviated as PDMS) molecules spreading 
on  bare oxidized silicon wafers, used without cleaning,
 or on wafers bearing loose
grafted trimethyl layers \cite{fraysse} demonstrate
that
$D_{exp}$ is with good accuracy
inversely proportional to the PDMS bulk viscosity $\eta(M)$.
Here, since for the
range of polymer weights studied in Ref.5 the 
best fit for the viscosity (Petrarch data for the PDMS) 
is given by the
algebraic law $\eta(M) \sim M^{1.7}$, the observations of Ref.5
appear to agree quite well with the 
results reported in  Refs.9 and 10.
We note also that the dependence 
$D_{exp} \sim 1/\eta(M)$ is consistent with the
theoretical prediction 
of the  hydrodynamical model of 
 mesoscopically thin 
films spreading  \cite{joann}. Such a model, however, is not justified
for films whose thickness amounts to only a few angstroms.
 
On the other hand, 
recent ellipsometric studies \cite{vil}
show that the $M$-dependence of $D_{exp}$ 
fades out at high RH
in case of PDMS molecules spreading 
on cleaned hydrophilic substrates. This
contradicts apparently to the results of 
Refs.9 and 10 and suggests 
that the exponent
$\gamma$ is not necessarily a constant, but
may vary with the RH.
As a matter of fact, the observation made in Ref.6 is in favor 
of the theoretical model 
of Ref.13, which predicts that dewetting dynamics of 
polymeric monolayers
is governed by the law in Eq.(1)
with $M$-independent $D_{exp}$.

In this Letter 
we report the results of first
$\em systematic$  analysis
of the $M$-dependence of $D_{exp}$, deduced from the experimental studies of  
rates of molecularly thin PDMS films spreading
on cleaned bare silicon wafers, 
exposed to an atmosphere with controllable, variable 
RH. 
Our experimental data
substantiate the $M$-dependence of $D_{exp}$ at different physical and chemical 
conditions. We
show that it is actually consistent
with the algebraic law in Eq.(2), but the exponent
$\gamma$ appears to be non-universal and 
depends, in general, on many different factors: 
For instance, it is very sensitive to
the chemical composition of the
solid surface, which  determines the frictional
properties of the surface with respect to the molecular transport.
Moreover, we find that $\gamma$ may  attain different values depending whether
only a single 
precursor film 
develops from the macroscopic droplet,
or in the terraced wetting case, 
when the precursor is formed by a succession 
of superimposed monolayers. In the former case, $\gamma$ is typically 
less than unity, while
for the latter case it may actually reach under certain conditions
the value $1.7$, as reported
previously in Refs.9 and 10.

Here we summarize the results obtained for 
the situation with a single precursor
film of monomolecular thickness.  
The parameters which are varied in our experimental 
studies are the polymer weight $M$
and the chemical composition 
of the substrate.
For the substrates used in this set of experiments, namely,
for the hydrophilic oxidized wafers, the frictional properties 
can be modified by changing the
RH. 
We perform our experiments at different RH and calculate the corresponding
 values of the
exponent $\gamma$. "Terraced" profiles are observed at very low RH ($\leq 20\%$),
while a single monolayer is observed for RH $\geq 35\%$. For such RH, 
which corresponds to
the limit of maximal friction in the single precursor film case,  
$\gamma$  is maximal, but  is substantially less than the value
reported in Refs.5,9 and 10. We find that here 
$\gamma \approx 1$.  
With an increase of RH, which lowers friction with the substrate,
$\gamma$ drops off monotonously
to zero. Note that $\gamma \approx 1$ is also found for compact 
hydrophobic grafted trimethyl layers, on
which the profile again exhibits a single step.
For two extreme cases of low ($\approx 35\%$) and very high
RH we propose simple arguments which 
explain the observed values $\gamma \approx 1$ and $\gamma \approx 0$.
Results of systematic experimental studies
of the behavior
in the terraced wetting case, 
for which  we actually recover
the value $\gamma \approx 1.7$ 
either at very low RH or on 
loose grafted trimethyl layers,
will be
presented in the complete paper \cite{vala}.

Ellipsometry is a well-known optical technique which 
is sensitive to the presence of films of molecular or
sub-molecular thickness, deposited at the solid surface
 with a different refractive index
\cite{el}. It thus allows to measure the thickness of a 
film and correspondingly, its radial extension,
by analyzing the change of polarization of an
optical beam at reflection. 

The present study was performed with a spatially resolved \cite{caza},
polarization
modulated ellipsometer working at a single wavelength
($6328$ $\AA$, He-Ne laser) and at the Brewster angle.
The setup has been extensively described elsewhere \cite{caza,fraysse}.
The liquids used in our experiments were 
low-$M$ 
polydimethylsiloxanes. Specifically, 
the PDMS are neutral, highly flexible
chain-like molecules with the transverse 
size of a monomer being approximately $7 \AA$; PDMS are 
liquids at ambient temperatures and 
glassify below $-123^{\circ}C$. The entanglement threshold $M_{c}$ 
in 3D
for PDMS is around $25000$ g/mol.
In our experiments, 
the  molecular weight (viscosity)
 range 
between $2000$ g/mol ($20$ cP) and $28400$ g/mol ($10^3$ cP).
  The polydispersity index is $1.7$ for the $20$ cP
(non fractionated oil) and between $1.05$ and $1.29$ for the
 heavier ones (fractionated) \cite{nathalie}. The molecular weight
distributions are monomodal, as checked by GPC \cite{caza}, which ensures that
the 
polydispersity effects are insignificant \cite{caza,fraysse,vil,nathalie}.
The substrates used in our measurements
were oxidized silicon wafers from electronic industry,
polished on one side and covered with a natural oxide. 
The wafers were cleaned by  
an oxygen flow under UV illumination, 
and then placed (for $24$ hours) 
in a measurement chamber with an
atmosphere with controlled RH. The purpose of the latter procedure
was to obtain on the wafer an equilibrated
 submonolayer (of controlled coverage, dependent on RH)
of water \cite{vila}. As we have already mentioned, 
the presence of
the water on top of our hydrophilic wafer is essential and allows us
to modify the frictional properties of the substrate
 by changing the chemical composition of the surface.

Experimentally obtained values of 
 $D_{exp}$ 
are summarized in Fig.1, 
where $D_{exp}$ is depicted as a function of the number of monomers $N$, which is
proportional to the molecular weight  $M$.
We present here four sets of data: Two sets are deduced from spreading rates of
the PDMS monolayers at low RH, (RH = $40 \%$), 
when only a very 
small portion of 
the surface is covered by  water molecules, which all
cluster in isolated, molecularly thin
 islands.  
Two other sets are
obtained from the data corresponding to the intermediate, (RH = $82 \%$), 
and high, (RH = $98 \%$),
relative humidities, when  water 
covers progressively higher and higher
areas of the surface. The data for the case RH = $98 \%$ 
shows that the $M$-dependence of
$D_{exp}$ fades out completely. As an illustration of such a behavior, 
 we depict in Fig.2 the dynamical thickness profiles
of different PDMS droplets, spreading on the same wafer
exposed to the atmosphere with RH = $98 \%$. Fig.2 shows 
that the radial extension of 
precursor films is absolutely independent of the molecular weight
at such high values of RH.

To explain the experimentally observed behavior we make
use of the so-called "stratified droplet" model, 
developed in Ref.7, and the microscopic
dynamical model of Ref.8. In these theoretical works 
it was shown that, in essence,
the behavior as in Eq.(1) results from the competition 
of two different factors -
the constant 
driving force, which stems from the presence of attractive liquid-solid interactions,
and  viscous-type dissipation for molecular motion on solid substrate; 
$D_{exp}$ is given by
\begin{equation}
D_{exp} \; \approx \; \frac{W_{2} \; - \; W_{1}}{\zeta},
\end{equation}
where $\zeta$ is the friction coefficient for molecular motion on solid substrate,
while the terms 
$W_{1}$ and $W_{2}$ describe the  energy of the liquid-solid and the liquid-liquid 
interactions of a molecule 
being directly on the top and 
at the height of one molecular size above the substrate.

For low-$M$ disentangled polymers, as it appears in our case, 
one evidently has that
\begin{equation}
W_{2} \; - \; W_{1} \; \sim \; M \; \triangle w,
\end{equation}
where $\triangle w$ is the corresponding gain of the  interaction 
energy due to moving of a monomeric unit 
from the second layer to the first one.
Consequently, the comparison between the experimentally 
observed behavior for low and high RH against 
 Eqs.(3) and (4) suggests that
\be \label{i}
\disp \zeta  \sim  \left\{ \begin{array}{ll}
M^2,  &  \mathrm{\rm for  \; low  \; RH, } \qquad  (a)\\
M, &  \mathrm{\rm for \;  high  \; RH,}  \qquad  (b)
\end{array}
\right.
\ee
which is reminiscent  
of the chain friction in melts (reptation)
and in the Rouse regime,  respectively \cite{pgga,pggb}.

Let us speculate now about the origin of the behavior described by Eq.(5.a).  A
 physically plausible explanation,
which parallels in many aspects
 the earlier discussion 
of slow spreading of polymer melts 
\cite{pggc,bruin} and slow desorption of polymer chains through an overlayer
of strongly adsorbed chains \cite{gran,grana},  may be as follows: 
The substrates used in our experimental studies, i.e. 
bare oxidized silicon
wafers, 
are chemically heterogeneous and contain different types of surface sites;
these are, namely,  
low energy siloxane bridges and high energy, chemically active silanol sites. 
The silanol  sites, which are present at a 
 density $n_{s} \approx 1$ $nm^{-2}$, 
(i.e. typically one such a site per each $10$ angstroms for our wafers),
may form a hydrogen bond with any of monomers of the 
PDMS molecules and thus represent the
"hot spots"  of  attachment to the surface.  
Consequently, the disentangled PDMS molecules, which emerge 
on the  wafer and form the precursor, 
get  trapped by the silanol sites.     
Since the density of these sites 
is rather
high, many of the PDMS monomers 
have a chance to form bonds, which means, first, that 
in accord with experimental observations
 the polymers will  not have
pronounced "loops" between the attached monomers and will 
lay almost flat on the substrate.  Second, it implies that
the effective time $\tau_{tr}(M)$, which a given PDMS 
molecule spends being trapped
by many silanol sites simultaneously, 
can be significantly larger than the life-time of a hydrogen bond
for a single monomer.

Now, following Refs.19 to 22, we suppose that 
in such a situation the transport of "mass" in the
 precursor
proceeds mainly by tortuous diffusion of a small portion of non-adsorbed
molecules, (for instance, those which emerge on the substrate 
already occupied by many other polymers), 
through a two-dimensional array of immobilized polymers, which serve for the latter
as obstacles. Motion of non-adsorbed macromolecules on solid surfaces
among immobilized, adsorbed polymers
was
first discussed in Ref.19. Assuming essentially reptative type of motion
of non-adsorbed
molecules, it has been shown that
apart of non-significant logarithmic factor $ln(M)$ resulting
from the presence of "loops" in adsorbed polymers, which is not the case here, 
the friction coefficient follows the dependence described by Eq.(5.a).
We  also remark that the behavior as in Eq.(5.a)
has been recently deduced from experimental studies of polymer adsorption on
substrates, grafted with other polymers \cite{grana}.  Lastly, 
we note that this reptation-like picture does not require 
the adsorbed chains to be irreversibly
trapped. However, the trapping time $\tau_{tr}(M)$ must be larger than 
some characteristic time involved in the non-adsorbed
chain dynamics (possibly, the time needed for one molecule 
to progress over its own length)
\cite{pggo}. 

Consider next how the situation can be changed, 
if prior to the deposition of a drop
the substrate was exposed to an atmosphere with
 high  RH. As a matter of fact,
the water molecules also have a strong 
affinity to the silanol sites and 
effectively screen  them, forming molecularly thin islands of water \cite{vila}.
Consequently, increasing the RH prior to the deposition of the droplet, 
we decrease substantially
the number of the silanol 
sites which can trap the PDMS molecules and 
thus make the substrate more
chemically homogeneous. On the other hand, the presence of interconnected
clusters of water, appearing at high RH, enhances the PDMS surface mobility
and facilitates mass transfer. 
Accordingly,
 at high RH one encounters an effectively ideal surface
with low friction, which seemingly 
explains
the Rouse-type behavior described by 
Eq.(5.b).

In conclusion, we have
presented here the results of systematic ellipsometric
studies 
of the polymer weight dependence of 
spreading rates of single monomolecular
 precursor films,
emitted by  droplets of low-$M$ 
polydimethylsiloxane molecules.
We have shown that 
the  "diffusion coefficient" $D_{exp}$ of the precursor
follows an algebraic dependence
on the molecular weight, $D_{exp} \sim M^{- \gamma}$,
where the exponent $\gamma$ depends on the frictional properties of 
the solid substrate. In case of hydrophilic substrates used in our studies,
such properties  are most conveniently 
 modified by the change of the RH.
Our experiments have shown that for high friction $\gamma$
is maximal, $\gamma \approx 1$, and that it tends to zero when
friction gets decreased.   
 For two extreme cases of high
 and very low friction
we propose  
simple arguments which 
 explain the experimentally observed
behavior.

\vspace{0.5cm} 

Enlightening discussions with P.G.de Gennes are gratefully
acknowledged. This research was supported by the European
Community with the grant CHRX-CT94-0448.


\end{document}